\documentstyle{article}

\catcode`\@=11
\def\hexnumber@#1{\ifcase#1 0\or1\or2\or3\or4\or5\or6\or7\or8\or9\or
	A\or B\or C\or D\or E\or F\fi }
\font\tenmsy=msym10

\font\sevenmsy=msym7

\font\fivemsy=msym5
\newfam\msyfam
\textfont\msyfam=\tenmsy  \scriptfont\msyfam=\sevenmsy
  \scriptscriptfont\msyfam=\fivemsy
\edef\msy@{\hexnumber@\msyfam}
\def\Bbb{\ifmmode\let\next\Bbb@\else
 \def\next{\errmessage{Use \string\Bbb\space only in math mode}}\fi\next}
\def\Bbb@#1{{\Bbb@@{#1}}}
\def\Bbb@@#1{\fam\msyfam#1}
\catcode`\@=12

\def\ie{\mbox{{\it i.e.} }}

\def\eg{\mbox{{\it e.g.} }}
\def\cf{\mbox{{\it cf.} }}
\def\etal{\mbox{{\it et al.} }}
\def\tr{\mbox{tr}}

\def\id{\mathop{\rm id}}
\def\field#1{{\Bbb #1}}
\def\real{{\field R}}
\def\complex{{\field C}}
\def\integer{{\field Z}}
\def\R{{\cal R}}

\def\matrix#1#2#3{#1 ^{#2}{}_{#3}}

\def\bracket#1#2{\left\langle #1\vphantom{#2}\right| \left.\vphantom{#1}#2
\right\rangle}

\let\vev\VEV
\def\abs#1{\left| #1\right|}
\def\half{{1\over 2}}
\def\inv#1{{1\over #1}}
\def\comm#1#2{\left[#1, #2\right]}
\def\inprod#1#2{\left\langle #1, #2\right\rangle}
\newfont{\gothic}{eufm10 scaled\magstep1}

\def\A{{\cal A}}
\def\G#1{\Gamma\left(#1\right)}
\def\M{{\cal M}}
\def\O#1#2{\Omega^{#1}\left(#2\right)}
\def\U{{\cal U}}
\def\uea#1#2{U_q(\mbox{\gothic #1}(#2))}
\def\DA{{\Delta_{\A}}}
\def\AD{{{}_{\A}\Delta}}
\def\ad{{\stackrel{\mbox{\scriptsize ad}}{\triangleright}}}
\def\metric#1#2{\eta^{(#1)}_{#2}}
\def\g{\mbox{\gothic g}}

\def\struc#1#2{f_{#1}{}^{#2}}
\def\gen#1{T_{#1}}
\def\quint#1{\left[#1\right]}
\def\adj{\mbox{\underline{ad}}}
\def\fun{\mbox{\underline{fn}}}
\def\trv{\mbox{\underline{tv}}}
\def\dg{\delta}
\def\dr{\mbox{d}}
\def\cov{\mbox{D}}
\def\bigA#1#2{\matrix{\field A}{#1}{#2}}
\def\Phibar{\Phi^{\dagger}}
\def\wein{\theta_{\rm W}}
\def\swein{\sin\wein}
\def\cwein{\cos\wein}
\def\twein{\tan\wein}
\def\dslash{\hbox{$\partial$\kern-1.2ex\raise.2ex\hbox{$/$}}}
\def\Dslash{\hbox{$\cov$\kern-1.5ex\raise.2ex\hbox{$/$}}}
\def\Wslash{\hbox{$W$\kern-1.6ex\raise.2ex\hbox{$/$}}}
\def\Aslash{\hbox{$A$\kern-1.3ex\raise.2ex\hbox{$/$}}}
\def\Zslash{\hbox{$Z$\kern-1.4ex\raise.2ex\hbox{$/$}}}
\def\Gslash{\hbox{$\Gamma$\kern-1.2ex\raise.2ex\hbox{$/$}}}

\catcode`\@=12
\catcode`\~=12

\begin{document}
\begin{titlepage}
\begin{center}
UMTG-189\hfill hep-th/9603143\\[.2in]

{\Large\bf Toward a $q$-Deformed Standard Model}\footnote{Research
supported in part by the National Science Foundation under grant
PHY-9507829}\\[.2in]
{\bf Paul Watts}\footnote{{\it watts@physics.miami.edu}, {\it
http://phyvax.ir.miami.edu:8001/watts/home.html}, (305)284-2325x8}\\[.1in]
{\it Department of Physics\footnote{Fax: (305)284-4222}\\
University of Miami\\
P.O.\ Box 248046\\
Coral Gables, FL 33124-8046\\
USA}\\[.2in]

\begin{abstract}

A gauge theory with an underlying $SU_q(2)$ quantum group symmetry is
introduced, and its properties examined.  With suitable assumptions, this
model is found to have many similarities with the usual $SU(2)\times U(1)$
Standard Model, specifically, the existence of four generators and thus
four gauge fields.  However, the two classical symmetries are unified into
one quantum symmetry, and therefore there is only a single coupling
constant, rather than two.  By incorporating a Higgs sector into the model,
one obtains several explicit tree-level predictions in the undeformed
limit, such as the Weinberg angle: $\sin^2\wein=\frac{3}{11}$.  With the
$Z$-boson mass $m_Z$ and fine structure constant $\alpha$ as inputs, one
can also obtain predictions for the weak coupling constant, the mass of the
$W$, and the Higgs VEV.  The breaking of the quantum invariance also
results in a remaining undeformed $U(1)$ gauge symmetry.

\end{abstract}
\end{center}
PACS-96: 12.10.Kt, 12.60.Cn, 12.60.Fr\\
MSC-91: 16W30
\end{titlepage}

\newpage

\setcounter{page}{1}
\renewcommand{\thepage}{\arabic{page}}
\setcounter{footnote}{0}
\renewcommand{\thefootnote}{\arabic{footnote}}

\tableofcontents


\section{Introduction}
\setcounter{equation}{0}

It has long been known that symmetries play a vital role in nature, and
that any physical theory describing observations must take these into
account.  This attitude has yielded amazing success, perhaps the most
notable example being the Standard Model (SM), in which it is assumed that
reality has a $SU(2)\times U(1)$ symmetry.

This symmetry group is a so-called ``classical'' group, \ie it satisfies
the usual properties of closure, existence of inverses and an identity, and
associativity.  There is, however, a more general notion which includes the
classical case, that of a matrix pseudogroup, or more familiarly, a quantum
group (QG) \cite{Drinfeld,Woronowicz3}.  This object is a ``deformed''
version of the classical case, in much the same way quantum mechanics is a
``deformed'' version of Newtonian mechanics (hence the moniker ``quantum
group'').  As such, one usually thinks of the deviation of a QG from the
usual case as parametrized by a number $q$, with $q=1$ giving the classical
case, just as $\hbar$ encapsulates the difference between classical and
quantum mechanics, with the correspondence principle allowing one to relate
the two.  Therefore, since the deformation of classical mechanics into
quantum mechanics results in new physics, it would not be surprising to
expect the same to occur in a theory with a QG symmetry.

In the SM, the symmetry group is taken to be local, and therefore must be
gauged, with the four particles needed to do this being identified with the
photon $A$ and the three weak-force mediators $W^{\pm}$ and $Z$.  However,
the overall symmetry is also assumed to be broken spontaneously via a Higgs
mechanism, leading to a unbroken $U(1)$ symmetry group and masses for the
weak bosons.

Since this approach is so remarkably successful in describing nature, if
one is interested in deforming the SM, it seems reasonable to incorporate
the same concepts into a theory with a QG symmetry.  This paper attempts to
do just that: To construct a gauge theory with a symmetry group
corresponding to the deformed version of $SU(2)$, denoted by $SU_q(2)$.
Not surprisingly, other authors have also pursued this line of thought
(with \cite{Cas1,Cas2} being particularly relevant to this work).  However,
this work approaches the problem via the deformation of the classical Lie
algebra {\gothic su}$(2)$ into the quantum Lie algebra (QLA)
\cite{Bernard,SWZ} $\uea{su}{2}$, and then uses the fact that $\uea{su}{2}$
has {\em four} generators, and therefore needs the same number of gauge
fields as the classical symmetry algebra {\gothic su}$(2)\oplus${\gothic
u}$(1)$ does.  Furthermore, there will be only {\em one} coupling constant
needed, since there is only one symmetry.

The approach presented here treads a thin line, perhaps.  It is necessary
to be somewhat abstract in order to discuss connections algebraically.  In
fact, this subject should technically be treated within the context of
sheaf theory: In the classical case (\cf \cite{Warner} \etal), this allows
topological properties of fiber bundles to be translated into algebraic
language, and there has been substantial progress in generalizing the same
basic theory to the deformed case as well \cite{Pflaum}.  (There have also
been some nice treatments of deformed gauge theories without resorting to
sheaves, \eg \cite{Brz}.)  However, since this work means to deal with more
physical aspects, the ultimate goal of interpreting these objects as gauge
fields is kept in mind, so some of the more mathematical details will be
deemphasized or ignored.

Finally, the author must stress that the model presented here is {\em
highly} speculative.  As the reader will see, there are plenty of
assumptions and leaps of faith made, some perhaps unjustifiably, in order
to obtain heuristic results.  In fact, it isn't clear that this particular
approach is even the best, since much work has already been done in the
subject by emphasizing the noncommutative geometric nature of the SM rather
than the QG aspect \cite{Lott,GnR}.  However, if one accepts the formal
existence of the objects appearing throught this work, then everything is
mathematically consistent.  Furthermore, all the predictions are given only
for the undeformed case anyway, where there are explicit and
well-understood expressions for everything (for example, the quadratic form
introduced in Section \ref{sec-action} simply becomes the usual integral at
$q=1$).  In any case, this model has some interesting features, and the
author's purpose in presenting the material herein is to introduce what may
perhaps be a starting point for further explorations.

A note: Although later sections will refer to some of the notations and
definitions in Sections \ref{def-gauge} and \ref{sec-suq}, it is not
absolutely necessary for the reader to spend an excessive amount of time on
these Sections, and s/he can move on to Section \ref{sec-Higgs} and, if
still interested, come back later for the mathematical details.

\section{Deformed Gauge Theories}\label{def-gauge}
\setcounter{equation}{0}

In order to generalize the concepts involved in discussing a deformed gauge
theory, one must use the language of Hopf algebras (HAs)
\cite{Sweedler,Abe,Majid1}, as is usually the case when talking about QGs.
As a result, this section will be rather abstract and mathematical.

Let $\M$ be a unital associative *-algebra and $\A$ a *-Hopf algebra (both
over a field $k$) which (left) coacts on $\M$ \cite{SWZ}, \ie there exists
a linear algebra map $\AD:\M\rightarrow\A\otimes\M$,
$\phi\mapsto\phi^{(1)'}\otimes\phi^{(2)}$, satisfying the following:
\begin{eqnarray}
\phi^{(1)'}\otimes\AD\left(\phi^{(2)}\right)&=&\Delta\left(\phi^{(1)'}
\right)\otimes\phi^{(2)},\nonumber\\
\epsilon\left(\phi^{(1)'}\right)\phi^{(2)}&=&\phi,\nonumber\\
\AD\left(\overline{\phi}\right)&=&\theta\left(\phi^{(1)'}\right)\otimes
\overline{\phi^{(2)}},\nonumber\\
\AD(1)&=&1\otimes 1,
\end{eqnarray}
where $\Delta$, $\epsilon$ and $\theta$ are respectively the coproduct,
counit, and involution on $\A$, and \={\mbox{ }} the involution on $\M$.
(A right coaction $\DA:\M\rightarrow\M\otimes\A$ may be defined
analogously.)

Now, suppose that $\dg$ is an exterior derivative on $\A$, and from this we
construct the universal differential calculus $(\O{}{\A},\dg)$, with
$\O{}{\A}$ being the universal differential envelope (UDE) of $\A$
\cite{connes,wor}.  This is given a $\integer_2$-graded HA structure in
accordance with \cite{SZ}.  We similarly assume that $\dr$ is an exterior
derivative on $\M$, and take $\O{}{\M}$ as the UDE (with the multiplication
in this algebra denoted by $\wedge$).  The coaction above extends to a map
on these UDEs, \ie $\AD:\O{p}{\M}\rightarrow\bigoplus_{
q=0}^p\O{p-q}{\A}\otimes\O{q}{\M}$, defined recursively by
\begin{equation}
\AD(\dr\phi)\equiv\dg\phi^{(1)'}\otimes\phi^{(2)}+(-1)^{\abs{\phi^{(1)'}}}
\phi^{(1)'}\otimes\dr\phi^{(2)},
\end{equation}
where $\abs{\phi}$ is the degree of $\phi$, \eg $p$ if $\phi\in\O{p}{\M}$.

\subsection{Connections, Field Strengths and Covariant Derivatives}

A connection on $\M$ is defined as a linear map $\Gamma:\O{p}{\A}
\rightarrow\O{p+1}{\M}$ satisfying the following:
\begin{eqnarray}
\G{1}&=&0;\nonumber\\
\G{\dg\alpha}&=&-\dr\G{\alpha};\nonumber\\
\AD(\G{\alpha})&=&(-1)^{\abs{\alpha_{(1)}}+\abs{\alpha_{(3)}}\left(\abs{
\alpha_{(2)}}+1\right)}\alpha_{(1)}S\left(\alpha_{(3)}\right)\otimes\G{
\alpha_{(2)}}\nonumber\\
&&-\dg\alpha_{(1)}S\left(\alpha_{(2)}\right)\otimes 1\label{connection}
\end{eqnarray}
($\alpha\in\O{}{\A}$).  This in turn allows the definition of the field
strength $F:\O{p}{\A}\rightarrow\O{p+2}{\M}$ via
\begin{equation}
F\left(\alpha\right):=\dr\G{\alpha}+(-1)^{\abs{\alpha_{(1)}}}\G{
\alpha_{(1)}}\wedge\G{\alpha_{(2)}}.
\end{equation}
The last of (\ref{connection}) then implies that
\begin{equation}
\AD(F(\alpha))=(-1)^{\abs{\alpha_{(2)}}\abs{\alpha_{(3)}}}\alpha_{(1)}
S\left(\alpha_{(3)}\right)\otimes F\left(\alpha_{(2)}\right).
\end{equation}

Let $\psi\in\O{p}{\M}$ be a $p$-form on $\M$; one can define a covariant
derivative $\cov$ as
\begin{equation}
\cov\psi:=\dr\psi+\G{\psi^{(1)'}}\wedge\psi^{(2)},
\end{equation}
so that $\cov$ maps $p$-forms into $(p+1)$-forms.  It follows immediately
from the coaction of $\A$ on $\Gamma$ that
\begin{equation}
\AD(\cov\psi)=(-1)^{\abs{\psi^{(1)'}}}\psi^{(1)'}\otimes\cov\psi^{(2)},
\end{equation}
which is exactly what one would want for a covariant derivative.  This, in
turn, gives $\cov^2\psi\equiv F\left(\psi^{(1)'}\right)\wedge\psi^{(2)}$,
as in the classical case.

To illustrate how this covariant derivative works, consider the following
example: Suppose that $\A$ is a QG, and $\psi^i$ a form living in the
fundamental rep of $\A$, so that if $\matrix{A}{i}{j}$ is the QG matrix
associated with this rep,
\begin{equation}
\AD\left(\psi^i\right):=\matrix{A}{i}{j}\otimes\psi^j;
\end{equation}
then if $\matrix{\Gamma}{i}{j}:=\G{\matrix{A}{i}{j}}$,
\begin{equation}
\AD\left(\matrix{\Gamma}{i}{j}\right)=\matrix{A}{i}{k}S\left(
\matrix{A}{\ell}{j}\right)\otimes\matrix{\Gamma}{k}{\ell}-\dg
\matrix{A}{i}{k}S\left(\matrix{A}{k}{j}\right)\otimes 1,
\end{equation}
and
\begin{equation}
\cov\psi^i\equiv\dr\psi^i+\matrix{\Gamma}{i}{j}\wedge\psi^j\mapsto
\matrix{A}{i}{j}\otimes\cov\psi^j.
\end{equation}

Notice that the field strength $\matrix{F}{i}{j}:=\dr\matrix{\Gamma}{i}{j}
+\matrix{\Gamma}{i}{k}\wedge\matrix{\Gamma}{k}{j}$ is thus a 2-form on $\M$
transforming according to
\begin{equation}
\AD\left(\matrix{F}{i}{j}\right)=\matrix{A}{i}{k}S\left(\matrix{A}{
\ell}{j}\right)\otimes\matrix{F}{k}{\ell}.
\end{equation}

Let $\U$ be the HA dually paired with $\A$; if $\A$ is a QG generated by
the matrix entries $\{\matrix{A}{i}{j}|i,j=1,\ldots,n\}$, then $\U$ is a
QLA generated by
\begin{equation}
\matrix{X}{i}{j}\equiv\gen{(ij)}:=\inv{\lambda}\inprod{1\otimes 1-\R_{21}
\R}{\matrix{A}{i}{j}\otimes\id}\label{def-X}
\end{equation}
where $\inprod{\mbox{ }}{\mbox{ }}$ is the dual pairing between $\U$ and
$\A$, $\R\in\U\otimes\U$ the universal R-matrix, and $\lambda$ is the
frequently occuring expression $q-q^{-1}$.  One can require that the
connection, when the argument is in $\O{0}{\A}\equiv\A$, takes the form
\begin{equation}
\G{a}\equiv\Gamma^A\inprod{\gen{A}}{a}=\Gamma^{(ij)}\inprod{\gen{(ij)}}{a}
\end{equation}
where $\Gamma^A\in\O{1}{\M}$ for $A=1,\ldots,n^2$.  The motivation for this
particular definition comes from the following two facts:  First, the
exterior derivative $\dg$ on $\A$ may be used to define a basis
$\{\omega^A\}$ for $\O{1}{\A}$ via
\begin{equation}
\dg a\equiv\omega^A\inprod{\gen{A}}{a_{(1)}}a_{(2)};
\end{equation}
secondly, with the quantum matrices $\bigA{A}{B}$
\begin{equation}
\bigA{(ij)}{(k\ell)}= S(\matrix{A}{k}{i})\matrix{A}{j}{\ell}
\end{equation}
giving the adjoint representation of the QLA via the right coaction of the
generators \cite{kill}, \ie
\begin{equation}
\DA\left(\gen{A}\right)=\gen{B}\otimes\bigA{B}{A},
\end{equation}
then it may be shown that by using $\AD\left(\Gamma(a)\right)=\AD\left(
\Gamma^A\right)\inprod{\gen{A}}{a}$, one obtains
\begin{equation}
\AD\left(\Gamma^A\right)\equiv\bigA{A}{B}\otimes\Gamma^B-\omega^A\otimes
1,\label{transf-gauge}
\end{equation}
the usual inhomogeneous transformation law for the components of $\Gamma$.

The above coaction will of course give the transformation properties of the
field strength; notice, however, that the definition of $F$ implies
\begin{equation}
F(a)=\dr\Gamma^A\inprod{\gen{A}}{a}+\Gamma^A\wedge\Gamma^B\inprod{\gen{A}
\gen{B}}{a}.
\end{equation}
Unlike the classical case, this does {\em not} necessarily have the form
$F^A\inprod{\gen{A}}{a}$; when the QLA is undeformed, the connection
1-forms anticommute, and $\gen{A}\gen{B}$ can be replaced by $\half\comm{
\gen{A}}{\gen{B}}$.  This is equivalent to $\half\struc{AB}{C}\gen{C}$, and
$F$ would be $\g$-valued.  Here, in the deformed situation, one must {\em
impose} appropriate anticommutation relations on the connections in order
for this to be the case.  If this is done, then the field strength is found
to transform homogeneously according to
\begin{equation}
\AD\left(F^A\right)=\bigA{A}{B}\otimes F^B,
\end{equation}
just as in the classical case.

\section{$SU_q(2)$}\label{sec-suq}
\setcounter{equation}{0}

Since the model presented here is considered to have a deformed $SU_q(2)$
symmetry, some details about this QG (and its associated QLA $\uea{su}{2}$)
must first be mentioned.

First of all, for any quasitriangular HA with a given universal
R-matrix, it is possible to construct the corresponding QLA
\cite{Jurco,Leipzig}.  $\uea{su}{2}$ is such a case \cite{Rosso}; however,
for calculational purposes, the more important object is the $4\times 4$
numerical R-matrix for the quantum group $SU_q(2)$; this is easily obtained
from the general form for the $SL_q(N)$ R-matrix \cite{FRT,linear} with
$N=2$, and with unitarity implying $q\in\real$:
\begin{equation}
R=q^{-\half}\left( \begin{array}{cccc}
q&0&0&0\\0&1&0&0\\0&\lambda &1&0\\0&0&0&q
\end{array}\right).\label{R-sl2}
\end{equation}
This is simply the universal R-matrix in the fundamental representation,
\ie $\matrix{R}{ij}{k\ell}:=\inprod{\R}{\matrix{U}{i}{k}\otimes\matrix{U}{
j}{\ell}}$, $U$ being the $2\times 2$ $SU_q(2)$ quantum matrix satisfying
$RU_1U_2=U_2U_1R$.  The $2\times 2$ matrix $X$ is given by (\ref{def-X})
with $A=U$, with the generators $\gen{1}$, $\gen{+}$, $\gen{-}$ and
$\gen{2}$ defined as
\begin{eqnarray}
X&:=&\inv{\lambda}\inprod{1\otimes 1-\R_{21}\R}{U\otimes\id}\nonumber\\
&=&\left(\begin{array}{cc}
\gen{1}&\gen{+}\\
\gen{-}&\gen{2}
\end{array}\right).\label{chi-fund}
\end{eqnarray}
The hermiticity condition which must be imposed on the generators of
$\uea{su}{N}$ is $\theta\left(\gen{(ij)}\right)=\gen{(ji)}$, and for $N=2$,
implies that $\gen{1}$ and $\gen{2}$ are self-adjoint, and $\theta
\left(\gen{\pm}\right)=\gen{\mp}$.

For the discussion presented here, a more convenient basis turns out to be
$\gen{0}=\gen{1}+\inv{q^2}\gen{2}$, $\gen{3}:=\frac{q^2}{1+q^2}(\gen{1}-
\gen{2})$, and $\gen{\pm}$.  Changing to these particular linear
combinations allows one to see more transparently the relation between the
deformed and undeformed cases.  For example, the QLA versions of the
adjoint action $\ad$ of $\uea{su}{2}$ on itself, written in terms of the
structure constants via $\gen{A}\ad\gen{B}=\struc{AB}{C}\gen{C}$
\cite{SWZ},
are
\begin{eqnarray}
\gen{0}\ad\gen{0}=0,&\gen{a}\ad\gen{0}=0,&\gen{0}\ad\gen{a}=-\lambda
\quint{2}\gen{a}\label{adj-basis1}
\end{eqnarray}
(where $a=\pm,3$), as well as
\begin{eqnarray}
\gen{3}\ad\gen{3}=-\lambda\gen{3},&&\gen{\pm}\ad\gen{\mp}=\pm\frac{
\quint{2}}{q}\gen{3},\nonumber\\
\gen{3}\ad\gen{\pm}=\pm q^{\mp 1}\gen{\pm},&&\gen{\pm}\ad\gen{3}=\mp q^{\pm
1}\gen{\pm},\label{adj-basis2}
\end{eqnarray}
where the ``quantum number'' $\quint{n}$ is defined as
\begin{equation}
\quint{n}:=\frac{1-q^{-2n}}{1-q^{-2}}.
\end{equation}
These of course reduce to the usual {\gothic su}$(2)$ commutators in the
$q\rightarrow 1$ limit.

The commutation relations themselves may also be found, and one finds that
$\gen{0}$ is central, and the other generators satisfy
\begin{eqnarray}
q^{\mp 1}\gen{3}\gen{\pm}-q^{\pm 1}\gen{\pm}\gen{3}&=&\pm\left( 1 -\frac{
\lambda}{\quint{2}}\gen{0}\right)\gen{\pm},\nonumber\\
\gen{+}\gen{-}-\gen{-}\gen{+}&=&\frac{\quint{2}}{q}\left(1 -
\frac{\lambda}{\quint{2}}\gen{0}\right)\gen{3}+\frac{\lambda\quint{2}}{q}
\gen{3}^2.\label{su2-comm}
\end{eqnarray}
There is also the identity
\begin{equation}
\left(1-\frac{\lambda}{\quint{2}}\gen{0}\right)^2=1+q^2\lambda^2J^2,
\label{det}
\end{equation}
where
\begin{equation}
J^2:=\inv{q^2\quint{2}}\left(q^2\gen{+}\gen{-}+\gen{-}\gen{+}+\quint{2}
\gen{3}^2 \right)\label{cas-ind}
\end{equation}
is the $\uea{su}{2}$ quadratic casimir.  Again, the classical limits of the
above give the familiar commutation relations.

The general definition of the $n\times n$ fundamental rep for a QLA is
\begin{equation}
\matrix{\fun}{i}{j}\left(\gen{(k\ell)}\right):=\inprod{\gen{(k\ell)}}{
\matrix{U}{i}{j}}=\inv{\lambda}\matrix{(I-R_{21}R)}{ki}{\ell j},
\end{equation}
so, by using (\ref{R-sl2}), the numerical matrices for the generators
$\{\gen{1},\gen{+},\gen{-},\gen{2}\}$ in the $2\times 2$ fundamental rep
$\fun$ of $\uea{su}{2}$ may be found, and when the change of basis above is
said and done, one gets
\begin{eqnarray}
\fun(\gen{0})=-\frac{\lambda}{q}\quint{\half}\quint{3\over 2}\left(
\begin{array}{cc}
1&0\\0&1\end{array}\right) ,&&
\fun(\gen{3})=\inv{\quint{2}}\left( \begin{array}{cc}
-1&0\\0&\inv{q^2}\end{array}\right) ,\nonumber\\
\fun(\gen{+})=\left( \begin{array}{cc}
0&0\\-\inv{q}&0\end{array}\right) ,&&\fun(\gen{-})=\left(
\begin{array}{cc}
0&-\inv{q}\\0&0\end{array}\right) .\label{fundrep}
\end{eqnarray}
(Note that $\gen{0}$ vanishes as $q\rightarrow 1$; this is what gets rid of
this extra generator for classical {\gothic su}$(2)$.)

The adjoint rep $\adj$ follows immediately from the adjoint actions given
in (\ref{adj-basis1}) and (\ref{adj-basis2}), and in the basis $\{\gen{0},
\gen{+},\gen{-},\gen{3}\}$, they take the forms
\begin{eqnarray}
\adj(\gen{0})=-\lambda\quint{2}\left( \begin{array}{cccc}
0&0&0&0\\0&1&0&0\\0&0&1&0\\0&0&0&1\end{array}\right),&&\adj(\gen{3})=
\left(\begin{array}{cccc}
0&0&0&0\\0&\inv{q}&0&0\\0&0&-q&0\\0&0&0&-\lambda\end{array}\right)
,\nonumber\\
\adj(\gen{+})=\left( \begin{array}{cccc}
0&0&0&0\\0&0&0&-q\quint{2}\\0&0&0&0\\0&0&\inv{q}&0\end{array}\right),&&
\adj(\gen{-})=\left( \begin{array}{cccc}
0&0&0&0\\0&0&0&0\\0&0&0&\frac{\quint{2}}{q}\\0&-\inv{q}&0&0\end{array}
\right).\label{adj-rep}
\end{eqnarray}

\section{A Deformed Standard Model}
\setcounter{equation}{0}

\subsection{$\uea{su}{2}$ Connection and Field Strengths}

Recall that in general $F(a)$ does not in general have the form $F^A
\inprod{\gen{A}}{a}$.  However, $F$ may be written in this way if one is
willing to {\em impose ad hoc} a certain set of anticommutation relations
on the connection 1-forms, which can be determined for the case where the
QLA being considered is $\uea{su}{2}$.  By using the generator commutation
relations (\ref{su2-comm}) and the identity (\ref{det}), it is
straightforward to see that the deformed anticommutation relations
\begin{eqnarray}
\Gamma^{\pm}\wedge\Gamma^{\pm}&=&0,\nonumber\\
\Gamma^{\pm}\wedge\Gamma^3+q^{\pm 2}\Gamma^3\wedge\Gamma^{\pm}&=&0,
\nonumber\\
\Gamma^{\pm}\wedge\Gamma^0+\Gamma^0\wedge\Gamma^{\pm}&=&\pm\frac{q^{\pm
1}\lambda}{\quint{2}}\Gamma^3\wedge\Gamma^{\pm},\nonumber\\
\Gamma^+\wedge\Gamma^-+\Gamma^-\wedge\Gamma^+&=&-q^2\quint{2}^2\Gamma^0
\wedge\Gamma^0,\nonumber\\
\Gamma^0\wedge\Gamma^3+\Gamma^3\wedge\Gamma^0&=&-\frac{\lambda}{q}\Gamma^-
\wedge\Gamma^+-\frac{\lambda\quint{2}}{q}\Gamma^0\wedge\Gamma^0,
\nonumber\\
\Gamma^3\wedge\Gamma^3&=&\frac{\lambda\quint{2}}{q}\Gamma^-\wedge\Gamma^+
-\frac{\quint{2}^2}{q^2}\Gamma^0\wedge\Gamma^0,\label{anticomm}
\end{eqnarray}
do indeed make $F$ $\uea{su}{2}$-valued, with components
\begin{eqnarray}
F^0&=&\dr\Gamma^0+\frac{2\quint{2}}{\lambda}\Gamma^0\wedge\Gamma^0,
\nonumber\\
F^{\pm}&=&\dr\Gamma^{\pm}\pm q^{\pm 1}\Gamma^3\wedge\Gamma^{\pm},
\nonumber\\
F^3&=&\dr\Gamma^3-\frac{\quint{2}}{q}\Gamma^-\wedge\Gamma^+-\frac{
\quint{2}^2}{q}\Gamma^0\wedge\Gamma^0.
\end{eqnarray}

All the above relations may be simplified with another ad hoc assumption:
$\Gamma^0\wedge\Gamma^0\equiv 0$.  This is certainly true in the undeformed
case, since $\Gamma^0$ is simply a 1-form; however, one must make sure that
this assumption is consistent with the deformed anticommutation relations
(\ref{anticomm}).  This is easily shown; using these relations, one finds
$\Gamma^0\wedge\Gamma^0$ commutes with each 1-form.  As for covariance
under the coaction $\AD$, notice from (\ref{transf-gauge}) that the
transformation for the connection component $\Gamma^0$ is
\begin{equation}
\AD\left(\Gamma^0\right)=1\otimes\Gamma^0-\omega^0\otimes 1.
\end{equation}
Since multiplication on the tensor product space $\O{}{\A}\otimes\O{}{\M}$
is $\integer_2$-graded, \ie
\begin{equation}
(\alpha\otimes\phi)(\beta\otimes\psi):=(-1)^{\abs{\phi}\abs{\beta}}\alpha
\beta\otimes\phi\psi,
\end{equation}
it immediately follows that
\begin{equation}
\AD\left(\Gamma^0\wedge\Gamma^0\right)=1\otimes\left(\Gamma^0\wedge
\Gamma^0\right)+\left(\omega^0\wedge\omega^0\right)\otimes 1.
\end{equation}
It was proven in \cite{linear} that $\omega^0$ is
nilpotent\footnote{$\omega^0=\inv{\quint{N}}\xi$ for $SU_q(N)$, where $\xi$
is the 1-form of \cite{linear}.}, so $\Gamma^0\wedge\Gamma^0$ is a
left-invariant 2-form.  Thus, the nilpotency of $\Gamma^0$ will be
preserved under the coaction of $\A$, so the anticommutation relations
(\ref{anticomm}) remain covariant even with this assumption, and this will
not break the $SU_q(2)$ symmetry.  (It should again be emphasized that this
nilpotency is an assumption, made purely to facilitate what follows.)

The commutation relations between the connections and their exterior
derivatives may be defined to be those such that the Bianchi identities
hold, \ie $\cov F^A\equiv\dr F^A+\struc{BC}{A}\Gamma^B\wedge F^C$ vanishes
identically.  This assumption is yet another one put in by hand, but has
the great advantage that the resulting relations are automatically
covariant.  These may be found in the Appendix, and lead to the following
commutation relations for the components of the field strength:
\begin{eqnarray}
F^3\wedge F^{\pm}-q^{\pm 2}F^{\pm}\wedge F^3&=&\pm q^{\pm 1}\lambda
\quint{2}F^0\wedge F^{\pm},\nonumber\\
F^+\wedge F^--F^-\wedge F^+&=&q\lambda F^0\wedge F^3+\frac{q\lambda}{
\quint{2}}F^3\wedge F^3,\nonumber\\
F^0\wedge F^a&=&F^a\wedge F^0.
\end{eqnarray}
(The closure of the algebra of the field strength components is a
consequence of general covariance; unlike the connection components, the
$F$s transform homogeneously in the adjoint rep.)

\bigskip

So far, nothing has been said about the hermiticity of the connection
components $\Gamma^0$, $\Gamma^{\pm}$ and $\Gamma^3$.  Since in physics one
likes to know about adjoints and conjugates, and these connections will
ultimately be identified with the $SU_q(2)$ gauge fields, it would be nice
to address this subject.

Note the following: At the level of the quantum group $SU_q(2)$, the
unitarity condition for the matrices $\matrix{U}{i}{j}$ in the fundamental
rep is given by $S(U)=U^{\dagger}=\theta\left(U^{\rm T}\right)$, just as in
the classical case (recall that $S(U)=U^{-1}$).  Therefore, if $\Gamma=
\Gamma^A\fun\left(\gen{A}\right)$ is the 1-form-valued $2\times 2$ matrix,
then note that by using the HA identity
$\inprod{\theta(x)}{a}\equiv\inprod{x}{(\theta\circ S)(a)}^*$ for $x\in\U$,
$a\in\A$, then
\begin{equation}
\Gamma^{\dagger}=\overline{\Gamma^{(ij)}}\fun\left(\gen{(ji)}\right).
\end{equation}
So if $\Gamma$ (and therefore the $2\times 2$ field strength matrix $F$) is
antihermitean, as is usually required, this implies that $\overline{
\Gamma^{(ij)}}=-\Gamma^{(ji)}$, so $\Gamma^0$ and $\Gamma^3$ are
antihermitean, and $\overline{\Gamma^{\pm}}=-\Gamma^{\mp}$, and,
consequently, the same for the field strength components.

\subsection{$SU_q(2)$ Yang-Mills Action}\label{sec-action}

In order to use what has been developed so far to build a physical theory,
more conditions must be met: First of all, assume the existence of a
quadratic form on the differential algebra $\O{}{\M}$, \ie a map
$\bracket{\mbox{ }}{\mbox{ }}:\O{}{\M}\otimes\O{}{\M}\rightarrow k$ (\eg
$\bracket{\phi}{\psi}:=\int_M\phi\wedge\star\psi$ for a Riemannian manifold
$M$, $\phi$ and $\psi$ $p$-forms on $M$, $k=\complex$ for $q=1$).
Furthermore, this form is understood to respect the coaction of $\A$ on
$\M$, \ie under $\AD$,
\begin{equation}
\bracket{\phi}{\psi}\mapsto\phi^{(1)'}\psi^{(1)'}\bracket{\phi^{(2)}}{
\psi^{(2)}}
\end{equation}
(so the quadratic form is not necessarily symmetric, since in general $\A$
is noncommutative).  Consistency with the involutions on $\A$, $\M$ and $k$
also requires
\begin{equation}
\bracket{\phi}{\psi}^*=\bracket{\bar{\psi}}{\bar{\phi}}.\label{inv-inn}
\end{equation}

The Killing metric for an arbitrary QLA was examined in detail in
\cite{kill}, and now enters into the picture.  Recall the definition: If
$\rho$ is a rep of a QLA with generators $\{\gen{A}|A=1,\ldots n^2\}$, then
the $n^2\times n^2$ numerical Killing metric is
\begin{equation}
\metric{\rho}{AB}:=\tr_{\rho}\left(u\gen{A}\gen{B}\right),
\end{equation}
with $u$ the element of $\U$ given in \cite{Drinfeld}.  As in the
undeformed case, this has certain invariance properties, which may be
written in terms of the adjoint quantum group matrices as
\begin{equation}
\metric{\rho}{CD}\bigA{C}{A}\bigA{D}{B}=\metric{\rho}{AB}1.
\end{equation}
Therefore, one notices that the quantity $\metric{\rho}{AB}\bracket{F^A}{
F^B}$ is left-invariant, and being quadratic in the field strength seems
like the perfect choice for the the gauge field kinetic energy term in the
action.

For the case of $\uea{su}{N}$, the Killing metric is block-diagonal in the
basis $\{\gen{0},\gen{a}|a=1,\ldots,N^2-1\}$, so if the field strength is
written as $F(a):=F^0\inprod{\gen{0}}{a}+F^a\inprod{\gen{a}}{a}$, the
Yang-Mills action (with coupling constant $\kappa$) takes the form
\begin{eqnarray}
S_{\mbox{\scriptsize YM}}&:=&-\inv{2\kappa^2}\metric{\rho}{AB}\bracket{
F^A}{F^B}\nonumber\\
&=&-\inv{2\kappa^2}\left\{\metric{\rho}{00}\bracket{F^0}{F^0}+\metric{
\rho}{ab}\bracket{F^a}{F^b}\right\}.\label{YM-action}
\end{eqnarray}

All of the above will hold for {\em any} rep $\rho$ of $\uea{su}{N}$.
However, for the specific case where one considers the adjoint rep of
$\uea{su}{2}$, where the generators are given by (\ref{adj-rep}), the
element $u$ and hence the $4\times 4$ Killing metric can be computed
explicitly, and in the basis $(0,+,-,3)$ are
\begin{eqnarray}
\adj(u)&=&\inv{q^4}\left( \begin{array}{cccc}
0&0&0&0\\0&q^2&0&0\\0&0&\inv{q^2}&0\\0&0&0&1\end{array}
\right),\nonumber\\
\metric{\adj}{AB}&=&\frac{\quint{4}}{q^3}\left( \begin{array}{cccc}
\frac{q\lambda^2\quint{2}^2\quint{3}}{\quint{4}}&0&0&0\\0&0&q&0\\
0&\inv{q}&0&0\\0&0&0&\frac{q}{\quint{2}}\end{array}\right).
\label{killing-adj}
\end{eqnarray}
Thus, (\ref{YM-action}) takes the form
\begin{eqnarray}
S_{\mbox{\scriptsize YM}}&=&-\frac{\quint{4}}{2\kappa^2q^2}\left\{
\bracket{F^+}{F^-}+\inv{q^2}\bracket{F^-}{F^+}+\inv{\quint{2}}\bracket{
F^3}{F^3}\right.\nonumber\\
&&\left.+\frac{\lambda^2\quint{2}^2\quint{3}}{\quint{4}}\bracket{F^0}{F^0}
\right\}.
\end{eqnarray}
(This is, of course, hermitean, due to (\ref{inv-inn}).)

This may be written in terms of the $\Gamma$s, of course, but in order to
connect with the undeformed case, define the four 1-forms $W^{\pm}$, $W^3$
and $B$ by
\begin{eqnarray}
\Gamma^{\pm}:=-\frac{ig\sqrt{2}}{\quint{2}}W^{\pm},&\Gamma^3:=-igW^3,
&\Gamma^0:=-\frac{ig}{\lambda}\sqrt{\frac{\quint{4}}{\quint{2}^3\quint{3}}}
B,\label{redef}
\end{eqnarray}
where
\begin{equation}
g:=q\kappa\sqrt{\frac{\quint{2}}{\quint{4}}}.
\end{equation}
From what was previously discussed about the antihermiticity of the
connections, it follows that $B$ and $W^3$ are self-adjoint and
$\overline{W^{\pm}}=W^{\mp}$.  Furthermore, the Yang-Mills action now takes
the form
\begin{eqnarray}
S_{\mbox{\scriptsize YM}}&=&\inv{\quint{2}}\bracket{\dr W^+}{\dr W^-}
+\inv{q^2\quint{2}}\bracket{\dr W^-}{\dr W^+}+\half\bracket{\dr
W^3}{\dr W^3}\nonumber\\
&&+\frac{ig}{q\quint{2}}\left(\bracket{\dr W^+}{W^3\wedge W^-}-\bracket{\dr
W^-}{W^3\wedge W^+}\right.\nonumber\\
&&+\bracket{\dr W^3}{W^-\wedge W^+}+\inv{q^2}\bracket{W^3\wedge W^-}{\dr
W^+}\nonumber\\
&&\left.-q^2\bracket{W^3\wedge W^+}{\dr W^-}+\bracket{W^-\wedge W^+}{\dr
W^3}\right)\nonumber\\
&&+\frac{g^2}{q\quint{2}}\left(\bracket{W^3\wedge W^+}{W^3\wedge W^-}+
\inv{q^2}\bracket{W^3\wedge W^-}{W^3\wedge W^+}\right.\nonumber\\
&&\left.-\frac{2}{q\quint{2}^2}\bracket{W^-\wedge W^+}{W^-\wedge
W^+}\right)+\half\bracket{\dr B}{\dr B},\label{YM}
\end{eqnarray}
Thus, in the $q\rightarrow 1$ limit, $S_{\mbox{\scriptsize YM}}$ is the
usual SM action for the gauge fields $W^{\pm}$, $W^3$ and $B$.

\subsection{Higgs Mechanism, Weinberg Angle, and Gauge Field
Masses}\label{sec-Higgs}

The Yang-Mills action (\ref{YM}) was explicitly constructed to be invariant
under the deformed symmetry group $SU_q(2)$, and bears a definite
resemblance to the YM term present in the $SU(2)\times U(1)$-symmetric SM.
In fact, in the $q\rightarrow 1$ limit, the two agree exactly.  However,
one might argue that this is not a profound result, since $SU_q(2)$ is
algebraically equivalent to $SU(2)\times U(1)$ anyway; nothing new is
really happening.

However, although the two groups are indeed the same at the {\em algebraic}
level, they are not at the {\em Hopf algebraic} level, due to the fact that
the additional structure (\eg the coproduct) mixes the ``$SU(2)$'' part
given by $\{\gen{\pm},\gen{3}\}$ and the central ``$U(1)$'' piece from
$\gen{0}$.  It is this mixing which changes the situation drastically: In
the undeformed SM, one is free to pick the normalizations of the $SU(2)$
and $U(1)$ gauge fields arbitrarily, since the overall symmetry is just the
product of the two groups.  However, in order to keep the {\em quantum}
symmetry, \ie the HA structure, intact, the relative sizes of the $1\times
1$ and $3\times 3$ pieces of the Killing metric $\metric{\rho}{00}$ and
$\metric{\rho}{ab}$ are fixed once the rep $\rho$ is picked.  The relative
normalizations of the $W$s and $B$ are therefore restricted if one requires
that this action becomes the familiar YM action in the classical limit.
Furthermore, the existence of one, rather than two, symmetry groups
explains why only the one coupling constant is present, rather than the two
appearing in the undeformed SM.

The consequences of this become manifest when one considers the coupling of
the gauge fields to matter.  Ordinarily, one picks this matter to live in a
specific rep of the symmetry group, and then finds the interactions with
the gauge fields via the covariant derivative, and the situation is no
different here.  All resulting matter-gauge interactions will depend only
on $g$.

To illustrate this, consider a complex matter doublet $\Phi^i\in\O{0}{\M}$
(and its conjugate $\Phibar_i:=\overline{\Phi^i}$) living in the fundamental
rep of $SU_q(2)$, \ie
\begin{eqnarray}
\Phi:=\left(\begin{array}{c}\phi^-\\ \phi^0\end{array}\right),&&\Phibar
:=\left(\begin{array}{cc}\phi^+&\bar{\phi}^0\end{array}\right).
\end{eqnarray}
Under the QG action, these transform respectively as
\begin{eqnarray}
\Phi^i\mapsto\matrix{U}{i}{j}\otimes\Phi^j,&&\Phibar_i
\mapsto S\left(\matrix{U}{j}{i}\right)\otimes\Phibar_j.
\end{eqnarray}
Not surprisingly, since the entries of $U$ do not commute, the requirement
that the commutation relations between the $\phi$s be covariant under the
above coactions implies that they too are deformed, and look like
\begin{eqnarray}
\phi^0\phi^{\pm}=\inv{q}\phi^{\pm}\phi^0,&&\bar{\phi}^0\phi^{\pm}=q
\phi^{\pm}\bar{\phi}^0,\nonumber\\
\phi^+\phi^-=\phi^-\phi^+,&&\bar{\phi}^0\phi^0=\phi^0\bar{\phi}^0-\frac{
\lambda}{q}\phi^+\phi^-.\label{Higgs-comm}
\end{eqnarray}
It follows that the quantity $\Phibar\Phi:=\overline{\Phi^i}\Phi^i\equiv
\bar{\phi}^0\phi^0+\phi^+\phi^-$ is central and invariant.  Therefore, the
appropriate kinetic energy term for this matter will be $\bracket{(\cov
\Phi)^{\dagger}}{\cov\Phi}$.  $\Phi$ lives in the fundamental, so its
covariant derivative is given by $\cov\Phi:=\dr\Phi+\Gamma^A\fun\left(
\gen{A}\right)\Phi$; using (\ref{fundrep}) and (\ref{redef}),
\begin{eqnarray}
\cov\phi^-&=&\dr\phi^-+\frac{ig}{q\quint{2}}\left(\sqrt{\frac{\quint{4}}{
\quint{2}\quint{3}}}\quint{\half}\quint{3\over 2}B+qW^3\right)\phi^-
\nonumber\\
&&+\frac{ig\sqrt{2}}{q\quint{2}}W^-\phi^0,\nonumber\\
\cov\phi^0&=&\dr\phi^0+\frac{ig}{q\quint{2}}\left(\sqrt{\frac{\quint{4}}{
\quint{2}\quint{3}}}\quint{\half}\quint{3\over 2}B-\inv{q}W^3\right)\phi^0
\nonumber\\
&&+\frac{ig\sqrt{2}}{q\quint{2}}W^+\phi^-.\label{cov-higgs}
\end{eqnarray}

Suppose there exists a map $V:\O{}{\M}\rightarrow k$ such that $V\left(
\Phibar\Phi\right)$ is hermitean, invariant, and bounded from below.  If
such a  $V$ exists, it plays the role of a potential, and
\begin{equation}
S_{\mbox{\scriptsize H}}=\bracket{(\cov\Phi)^{\dagger}}{\cov\Phi}-V\left(
\Phibar\Phi\right)
\end{equation}
is an invariant hermitean action for $\Phi$.  Even though this action is
manifestly $SU_q(2)$-invariant, assume that the quantum symmetry is broken
spontaneously via a Higgs mechanism.  This is accomplished by assuming that
there is a real nonzero constant $v$ such that $V$ is minimized (and
vanishes) at $\half v^21$ (where $1$ is the unit in $\O{}{\M}$, which will
be suppressed from now on).  Therefore, the vacuum for the action above
occurs at $\vev{\Phibar\Phi}=\half v^2$, which one can assume corresponds
to $\vev{\phi^{\pm}}=0$ and $\vev{\phi^0}=\vev{\bar{\phi}^0}=\inv{\sqrt{2}}
v$.  (An example of a map satisfying all these conditions might be
something like
\begin{equation}
V\left(\Phibar\Phi\right):=\frac{\mu^2}{v^2}\bracket{\Phibar\Phi-\half
v^2}{\Phibar\Phi-\half v^2},
\end{equation}
with $\mu\in\real$.)

If all the above is possible, then the action will acquire terms quadratic
in the gauge fields, and thus they will become massive.  Now, just as in
the classical case, assume that the mass eigenstates are $W^{\pm}$, $Z$ and
$A$, where
\begin{eqnarray}
W^3\equiv\cwein Z+\swein A,&&B\equiv-\swein Z+\cwein A,
\end{eqnarray}
where $\wein\in\real$ is the Weinberg angle.  However, in order for $A$ to
be interpretable as the photon (more on this in the next Subsection), it
must be massless, which implies that $\cov\Phi$ cannot include a term of
the form $A\phi^0$ in $\cov\phi^0$, since $\vev{\phi^0}\neq 0$.  By using
the explicit form (\ref{cov-higgs}) and the definitions of $Z$ and $A$
above, this may be accomplished by requiring
\begin{equation}
\twein=q\sqrt{\frac{\quint{4}}{\quint{2}\quint{3}}}\quint{\half}\quint{
3\over 2}.
\end{equation}

Note two things: First of all, this expression for $\twein$ is independent
of the coupling constant $g$, unlike the classical case where it is given
as the ratio of the $SU(2)$ and $U(1)$ couplings $g$ and $g'$.  Secondly,
even though $\twein$ is still a function of $q$, if one assumes that the
``real world'' lives at (or at least very close to) $q=1$, s/he therefore
concludes that $\twein=\sqrt{\frac{3}{8}}$, or alternatively, $\sin^2\wein
=\frac{3}{11}=0.273$.  The experimental value is $0.2319$ \cite{PDB}, so
the predicted value is within 20\%.  (Recall that this is a tree-level
prediction only; no mention has been made of quantum effects.)

By inserting the value for $\wein$ from the above relation, the covariant
derivatives take the form
\begin{eqnarray}
\cov\phi^-&=&\dr\phi^-+\frac{ig}{\cwein}\left(\inv{\quint{2}}-\sin^2
\wein\right)Z\phi^-+\frac{ig\sqrt{2}}{q\quint{2}}W^-\phi^0+ig\swein
A\phi^-,\nonumber\\
\cov\phi^0&=&\dr\phi^0-\frac{ig}{q^2\quint{2}\cwein}Z\phi^0+\frac{ig
\sqrt{2}}{q\quint{2}}W^+\phi^-.\label{cov-Higgs}
\end{eqnarray}
An immediate result of this is that $\phi^-$ has electric charge
$-g\swein$.  Assuming that this is equal to the charge of the electron,
then by using the value of the fine structure constant $\alpha^{-1}=
137.04$ \cite{PDB}, one finds $g=\frac{e}{\swein}=\sqrt{\frac{44\pi
\alpha}{3}}=0.580$.

The masses of the three remaining gauge fields are found by evaluating
$S_{\mbox{\scriptsize H}}$ at $\vev{\Phi}$, giving
\begin{equation}
S_{\mbox{\scriptsize H}}|_{\vev{\Phi}}=\frac{g^2v^2}{q^2\quint{2}^2}
\bracket{W^+}{W^-}+\frac{g^2v^2}{2q^4\quint{2}^2\cos^2\wein}\bracket{Z}{Z}.
\end{equation}
Requiring this to be equal to $m_W^2\bracket{W^+}{W^-}+\half m_Z^2\bracket{
Z}{Z}$ determines the masses:
\begin{equation}
m_W=\frac{gv}{q\quint{2}}=qm_Z\cwein,
\end{equation}
which gives $m_W\rightarrow m_Z\cwein$ in the classical case, so if $m_Z
=91.187$ GeV \cite{PDB}, $m_W=m_Z\cwein=\sqrt{\frac{8}{11}}m_Z= 77.76$ GeV.
The experimental value is $m_W=80.22$ GeV, which is about 3\% away.
Furthermore, the Higgs VEV is given by $v=\frac{q^2\quint{2}\sin
2\wein}{2\sqrt{4\pi\alpha}}m_Z$, which, after taking $q\rightarrow 1$ and
sticking all the numbers in, is $v=268.21$ GeV.

\subsection{$U(1)$ Symmetry and Electric Charges}

To address the question of whether or not there remains any symmetry after
$\Phi$ gets a nonzero VEV, one can introduce two new fields $H$ and $\phi$
defined by
\begin{eqnarray}
H:=\sqrt{2}\quint{\half}\left(\bar{\phi}^0+\inv{q}\phi^0\right)-v,&&\phi
:=\frac{\sqrt{2}}{iq}\quint{\half}\left(\phi^0-\bar{\phi}^0\right),
\end{eqnarray}
both of which have vanishing VEVs.  Reexpressing (\ref{Higgs-comm}) in
terms of these new fields gives
\begin{eqnarray}
H\phi^{\pm}&=&\phi^{\pm}H+i(1-q)\phi^{\pm}\phi,\nonumber\\
\phi\phi^{\pm}&=&\left(q+\inv{q}-1\right)\phi^{\pm}\phi+i\left(1-\inv{q}
\right)\phi^{\pm}H+i\left(1-\inv{q}\right)v\phi^{\pm},\nonumber\\
H\phi&=&\phi H+2i\left(1-\inv{q}\right)\phi^+\phi^-.
\end{eqnarray}

Notice that the last term in the second of these is linear in $\phi^{\pm}$,
whereas every other term is quadratic in the fields; this is what breaks
the $SU_q(2)$ symmetry.  In fact, the only coactions which preserve these
commutation relations and are linear in the fields are
\begin{eqnarray}
H\mapsto 1\otimes H,&\phi\mapsto 1\otimes\phi,&\phi^{\pm}\mapsto a^{\pm
1}\otimes\phi^{\pm},
\end{eqnarray}
where $a$ is the sole generator of a QG with the relations $\Delta(a)=a
\otimes a$, $\epsilon(a)=1$, and $S(a)=\theta(a)=a^{-1}$.  Note that this
QG is entirely abelian, and thus is equivalent to the {\em classical} group
$U(1)$.

One may define a new derivative $\cov'$ by subtracting off the VEV of
$\phi^0$ from (\ref{cov-Higgs}), \ie
\begin{eqnarray}
\cov'\phi^-&:=&\cov\phi^--\frac{igv}{q\quint{2}}W^-,\nonumber\\
\cov'\left(\phi^0-\inv{\sqrt{2}}v\right)&:=&\cov\phi^0+\frac{igv}{q^2
\sqrt{2}\quint{2}\cwein}Z.
\end{eqnarray}
Under the remaining $U(1)$ symmetry, $\cov'$ is in fact a covariant
derivative, provided the gauge fields transform according to
\begin{eqnarray}
W^{\pm}\mapsto e^{\pm ig\swein\chi}\otimes W^{\pm},&Z\mapsto 1\otimes Z,&A
\mapsto 1\otimes A+\dg\chi\otimes 1,
\end{eqnarray}
where $a:=e^{ig\swein\chi}$.  These are precisely the gauge transformations
for a classical gauged $U(1)$, so the interpretation of $A$ as the photon
is indeed justified after all.

At the QLA, rather than the QG, level, the nonvanishing of $\vev{\phi^0}$
means that the vacuum state $\vev{\Phi}$ is no longer null, and that one
should look for all $2\times 2$ matrices annihilating this state in order
to see what remains after the spontaneous symmetry breaking.  This is
straightforward; all matrices proportional to $diag(-1,0)$.  Now, note that
if a quantum group $\A$ left coacts on a field $\psi$, then the dual QLA
$\U$ is linearly right represented on this same field via
\begin{equation}
\psi\triangleleft x:=\inprod{x}{\psi^{(1)'}}\psi^{(2)}
\end{equation}
for any element $x\in\U$ (so that $\phi\triangleleft(xy)=(\phi
\triangleleft x)\triangleleft y$).  Therefore, in the case being
considered, an element of $\uea{su}{2}$ acts on $\Phi$ as
\begin{equation}
\Phi^i\triangleleft x=\inprod{x}{\matrix{U}{i}{j}}\Phi^j=\matrix{
\fun}{i}{j}\left(x\right)\Phi^j.
\end{equation}
If $Q$ is the element of $\uea{su}{2}$ which is equal to $diag(-1,0)$ in
the fundamental, then $Q\equiv\frac{q}{\lambda\quint{2}\quint{\half}
\quint{3\over 2}}\gen{0}+\gen{3}$.  This is central, so the remaining
symmetry subalgebra is abelian, and is thus the classical algebra {\gothic
u}(1).

This can then be gauged, with the action of $Q$ on a field $\psi$ living in
rep $\rho$ given by $\psi\triangleleft Q=\rho(Q)\psi$, and when $\gen{0}$
is eliminated in favor of $\gen{3}$ and $Q$, then the covariant derivative
$\cov'$ in rep $\rho$ takes the form
\begin{eqnarray}
\cov'\psi&=&\dr\psi-\frac{ig\sqrt{2}}{\quint{2}}\left(W^+\rho\left(\gen{+}
\right)+W^-\rho\left(\gen{-}\right)\right)\psi\nonumber\\
&&-\frac{ig}{\cwein}Z\left(\rho\left(\gen{3}\right)-\sin^2\wein\rho(Q)
\right)\psi-ig\swein A\rho(Q)\psi.
\end{eqnarray}

\subsection{Fermions}

The undeformed SM includes fermionic matter, of course, so now an attempt
is made to put the same into the deformed model considered herein.  To this
end, let $\Psi^i$ be a fermion doublet living in the fundamental, with
$\bar{\Psi}_i$ its adjoint; in components,
\begin{eqnarray}
\Psi:=\left(\begin{array}{c}\psi\\\nu\end{array}\right),&&\bar{\Psi}:=
\left(\begin{array}{cc}\bar{\psi}&\bar{\nu}\end{array}\right).
\label{fermion}
\end{eqnarray}

Assume that there is a covariant derivative $\Dslash'$ for fermions
which preserves the transformation properties (which in the familiar
undeformed case is the usual $\Dslash':=\gamma^{\mu}\cov_{\mu}'$, but
here it is just assumed to exist without specifying its particular form).
Then, if the contribution to the total action is taken to be $S_{\rm
F}:=\bracket{\bar{\Psi}}{i\Dslash'\Psi}$, one finds 
\begin{eqnarray}
S_{\rm F}&=&\bracket{\bar{\psi}}{i\dslash\psi}+\bracket{\bar{
\nu}}{i\dslash\nu}\nonumber\\
&&-g\swein\bracket{\bar{\psi}}{\Aslash\psi}-\frac{g\sqrt{2}}{q\quint{2}}
\left(\bracket{\bar{\psi}}{\Wslash^-\nu}+\bracket{\bar{\nu}}{\Wslash^+
\psi}\right)\nonumber\\
&&+\frac{g}{\cwein}\left(\left(-\inv{\quint{2}}+\sin^2\wein\right)\bracket{
\bar{\psi}}{\Zslash\psi}+\inv{q^2\quint{2}}\bracket{\bar{\nu}}{\Zslash
\nu}\right).\label{ferm-left}
\end{eqnarray}
So, $\psi$ has electric charge $-g\swein$ and $\nu$ is neutral.

In principle, the $W-\nu-\psi$ coupling will result in a four-fermion
interaction in the low-energy theory, as in the classical case, and thus
would give the Fermi coupling constant $G_{\rm F}$.  In the $q\rightarrow
1$ limit, by using the value of $g$ from before, this gives a prediction of
$G_{\rm F}:=\frac{g^2}{4\sqrt{2}m_W^2}=\frac{121\pi\alpha}{24\sqrt{2}m_Z^2}
=0.983\times 10^{-5}$ GeV$^{-2}$, about 16\% away from the value in
\cite{PDB}, $1.16639\times 10^{-5}$ GeV$^{-2}$.

\section{Conclusions}
\setcounter{equation}{0}

This model, despite some interesting features, is still very minimal; it
cannot be considered a truly deformed version of the SM as it stands.
There are still several areas which must be looked into if there is any
hope whatsoever of treating the results presented above seriously.  Some of
these concerns will now be briefly addressed.

\subsection{Mathematical Concerns}

The actual geometrical interpretation of non(anti)commuting differential
forms as introduced here is a bit unclear.  Suppose one does in fact have
the four connection 1-forms satisfying the relations (\ref{anticomm});
classically, it is natural to immediately write $\Gamma^A:=\Gamma^A_{\mu}
\dr x^{\mu}$, where the $\{x^{\mu}\}$ are a set of local coordinates on the
space-time considered.  However, how is the noncommutative nature of the
connections manifested?  Do the individual components $\{\Gamma^A_{\mu}\}$
commute, but the coordinates do not, as in the case of the so-called Manin
plane \cite{Manin}?  Or is the space-time classical and the components form
some nontrivial algebra \cite{Sudbery}?  Or both?  Or neither?  Or can one
even hope to interpret the connections in this way?

Then there is the question of what this mysterious quadratic form
$\bracket{\mbox{ }}{\mbox{ }}$ actually is; in the classical case, as
already mentioned, it is just an integration over the space-time.  To
extend this to the QG case, one needs to understand how to integrate over
noncommuting objects.  In the case of strictly {\em anticommuting}
variables, this was done long ago with the Berezin integral \cite{Berezin},
and there has also been extensive work done for the general noncommuting
situation (\cf \cite{Chryss} and references therein).  Thus, it seems like
the possibility of building actions using these latter types of integrals
may exist, as it did in the case of supersymmetry using the former.

\subsection{Physical Concerns}

\subsubsection{Chiral Matter}

One obvious physical shortcoming of this model as a truly deformed version
of the SM is the fact that it seems to preclude the inclusion of chiral
fermions.  After all, this is one of the reasons the SM has a $SU(2)\times
U(1)$ symmetry, as opposed to something like a $U(2)$ symmetry: The
left-handed fermions live in the $(\half,0)$ rep and the right-handed ones
in the $(0,Y)$ rep ($Y$ being the $U(1)$ hypercharge of the fermion).  In
the model herein, it looks like the only possible way to incorporate
chirality would be to have the left-handed fermions in a $SU_q(2)$ doublet
and the right-handed ones in a singlet, \ie the trivial rep. Unfortunately,
it would seem that this is problematic, because the trivial rep is given by
the vanishing of all the generators, and would therefore give no coupling
of the right-handed fermions to the gauge fields.

However, all may not be lost; it is true that {\em a} 1-dimensional rep of
$SU_q(2)$ has all the generators vanishing.  But recall (\ref{su2-comm})
and (\ref{det}): These relations are also satisfied by the instance where
$\gen{\pm}$ and $\gen{3}$ vanish, but $\gen{0}$ is represented by
$\frac{2\quint{2}}{\lambda}$.  Thus, if $\chi$ is a fermion living in this
``trivial'' rep $\trv$, its contribution to the action may be taken to be
\begin{eqnarray}
\bracket{\bar{\chi}}{i\trv\left(\Dslash'\right)\chi}&=&\bracket{\bar{
\chi}}{i\dslash\chi}+\frac{2i\quint{2}}{\lambda}\bracket{\bar{
\chi}}{\Gslash^0\chi}\nonumber\\
&=&\bracket{\bar{\chi}}{i\dslash\chi}-\frac{g}{\cwein}\left(
\frac{2\sin^2\wein}{q\lambda^2\quint{\half}\quint{3\over 2}}\right)
\bracket{\bar{\chi}}{\Zslash\chi}\nonumber\\
&&+g\swein\left(\frac{2}{q\lambda^2\quint{\half}\quint{3\over 2}}\right)
\bracket{\bar{\chi}}{\Aslash\chi}.\label{ferm-right}
\end{eqnarray}
So this assumption does indeed allow one to couple the fermion with the $Z$
and the photon {\em without} coupling it to $W^{\pm}$, exactly what one
would want for a right-handed fermion.

However, there is still the matter of combining such a fermion with a
doublet; the obvious thing to try would be to take the singlet and doublet
fermions given by
\begin{eqnarray}
\chi:=\half\left(1+\gamma_5\right)\psi,&&\Psi:=\half\left(1-\gamma_5\right)
\left(\begin{array}{c}\psi\\\nu\end{array}\right).
\end{eqnarray}
Then, one would want to manipulate the kinetic energy terms for each of
these (given by (\ref{ferm-left}) and (\ref{ferm-right})) and collect all
the terms quadratic in $\psi$ together in such a way that the $Z$ and
photon couplings work out (the $W^{\pm}$ coupling between $\psi$ and $\nu$
already have the correct form).  The author has so far been unsuccessful in
this; perhaps some readers of this paper may be able to accomplish this
task.

\subsubsection{Quarks}

Even if the chiral problem can be solved, one must still try to incorporate
all the observed matter into the model, which must include quarks.  As in
the classical case, these must have an $SU(3)$ color symmetry (which would
remain undeformed, presumably), but also have to have appropriate $SU_q(2)$
properties.  There may already be some hint as to what these properties
might be: Remember (\ref{cas-ind}), where there is an explicit relation
between the {\gothic u}$(1)$ charge $Q$, the ``$z$-spin component''
$\gen{3}$ and the casimir $J^2$.  In particular, in the $q\rightarrow 1$
limit, $J^2=\frac{3}{2}\left(\gen{3}-Q\right)$, so that a state in the
spin-$j$ irrep of $SU(2)$ with $z$-component quantum number $m$ has
\begin{equation}
Q_{j,m}=m-\frac{2}{3}j(j+1).
\end{equation}
(This replaces the classical Gell-Mann-Nishijima relation $Q=\gen{3}+\half
Y$.)  For $j=\half$, this gives charges of $0$ for the $m=+\half$ state and
$-1$ for the $m=-\half$ state, as desired.  But what about other reps?  The
next one to consider is the adjoint, \ie $j=1$, for which $Q_{1,m}=m-\frac{
4}{3}$.  For $m=(+1,0,-1)$, the charges are $\left(-\frac{1}{3},-\frac{
4}{3},-\frac{7}{3}\right)$.  Even though no known multiplet has such
charges, the appearance of the 3 in the denominators is intriguing, and
indicates that perhaps there is some way of putting quarks into the theory
via the adjoint rep.

\subsubsection{Higgs Potential and Couplings}

The Higgs potential $V\left(\Phibar\Phi\right)$ was just included with the
assumption that it has a minimum which will give a nonzero VEV to $\Phi$,
without anything being said about its actual form (although an example was
given).  This unfortunately looks like it would have to be put in by hand
in this model, unlike in \cite{Lott}, where it arises automatically as a
consequence of the noncommutative nature of the model.

How about couplings between the Higgs and fermions?  One might initially
try to find invariant combinations of the Higgs doublet and a fermion
doublet $\Psi$ as in (\ref{fermion}).  One springs immediately to mind: The
obvious one given by $\bar{\Psi}\Phi\equiv\bar{\psi}\phi^-+\bar{\nu}
\phi^0$.  Another can be constructed following the classical example, using
the deformed Levi-Civita symbol for $SU_q(2)$, denoted by
$\epsilon^{ij}_q$.  This object has the values
\begin{eqnarray}
\epsilon^{12}_q=q^{\half},&\epsilon^{21}_q=-q^{-\half},&\epsilon^{11}_q=
\epsilon^{22}_q=0,
\end{eqnarray}
so that the unit determinant condition on the $2\times 2$ $SU_q(2)$
matrices can then be written as
\begin{equation}
\epsilon^{k\ell}_q\matrix{U}{i}{k}\matrix{U}{j}{\ell}\equiv\epsilon^{ij
}_q1.
\end{equation}
This means that one can define the doublet $\tilde{\Phi}^i$ by
\begin{equation}
\tilde{\Phi}^i:=\epsilon^{ji}_q\Phibar_j=\left(\begin{array}{c}
-q^{\half}\bar{\phi}^0\\q^{-\half}\phi^+\end{array}\right).
\end{equation}
Under the action of the QG, this transforms as $\AD\left(\tilde{\Phi}^i
\right)=\matrix{U}{i}{j}\otimes\tilde{\Phi}^j$, so the combination $\bar{
\Psi}\tilde{\Phi}$ is also invariant under the QG.

Classically, one would then take one of these two combinations and
construct a charge-conserving interaction term by using a fermion of
opposite chirality.  For example, if $\chi$ is the right-handed partner of
$\psi$, then
\begin{equation}
\bracket{\bar{\Psi}\tilde{\Phi}}{\chi}=q^{-\half}\bracket{\bar{\nu}\phi^+}{ 
\chi}-q^{\half}\bracket{\bar{\psi}\bar{\phi}^0}{\chi}
\end{equation}
would seem to be a possible Yukawa coupling between the chiral fermions and
the Higgs.  Unfortunately, if $\chi$ is thought of as a fermion in the
``trivial'' rep suggested above, then this term is {\em not} $SU_q(2)$
invariant, since $\gen{0}$ is nonzero in this rep.  One actually wants a
combination of $\bar{\Psi}$ and $\Phi$ which also lives in this ``trivial''
rep, and transforms in such a way so as to cancel out the variation of
$\chi$.  The author has yet to find such a combination, so the coupling of
the Higgs to any chiral fermions is still absent in this model.

\section*{Acknowledgements}

I would like to thank Markus Pflaum for discussions on the material in
Section \ref{def-gauge}, and Orlando Alvarez and Raphael Nepomechie for
reading the manuscript and offering several helpful suggestions.  I am also
deeply indebted to Bruno Zumino for invaluable help and inspiration.

\noindent This research was supported in part by the National Science
Foundation under grant PHY-9507826.
\newpage
\addcontentsline{toc}{section}{Appendix:  $SU_q(2)$ Connection Commutation
Relations}
\section*{Appendix:  $SU_q(2)$ Connection Commutation Relations}
\subsection*{Connection 1-Forms}
\begin{eqnarray*}
\Gamma^{\pm}\wedge\Gamma^{\pm}&=&0\\
\Gamma^{\pm}\wedge\Gamma^3+q^{\pm 2}\Gamma^3\wedge\Gamma^{\pm}&=&0\\
\Gamma^{\pm}\wedge\Gamma^0+\Gamma^0\wedge\Gamma^{\pm}&=&\pm\frac{q^{\pm
1}\lambda}{\quint{2}}\Gamma^3\wedge\Gamma^{\pm}\\
\Gamma^+\wedge\Gamma^-+\Gamma^-\wedge\Gamma^+&=&-q^2\quint{2}^2\Gamma^0
\wedge\Gamma^0\\
\Gamma^0\wedge\Gamma^3+\Gamma^3\wedge\Gamma^0&=&-\frac{\lambda}{q}\Gamma^-
\wedge\Gamma^+-\frac{\lambda\quint{2}}{q}\Gamma^0\wedge\Gamma^0\\
\Gamma^3\wedge\Gamma^3&=&\frac{\lambda\quint{2}}{q}\Gamma^-\wedge\Gamma^+
-\frac{\quint{2}^2}{q^2}\Gamma^0\wedge\Gamma^0
\end{eqnarray*}
\subsection*{$\dr\Gamma$ Commutation Relations}
\begin{eqnarray*}
\dr\Gamma^0\wedge\Gamma^a&=&\Gamma^a\wedge\dr\Gamma^0\\
\dr\Gamma^{\pm}\wedge\Gamma^{\pm}-\Gamma^{\pm}\wedge\dr\Gamma^{\pm}&=&0\\
\dr\Gamma^{\pm}\wedge\Gamma^{\mp}-\Gamma^{\mp}\wedge\dr\Gamma^{\pm}&=&
\pm q\lambda\Gamma^0\wedge\dr\Gamma^3\pm\frac{q\lambda}{\quint{2}}\Gamma^3
\wedge\dr\Gamma^3\\
&&\mp\lambda\quint{2}\Gamma^0\wedge\Gamma^-\wedge\Gamma^+\\
\dr\Gamma^{\pm}\wedge\Gamma^3-\Gamma^3\wedge\dr\Gamma^{\pm}&=&\mp q^{\pm
1}\lambda\Gamma^{\pm}\wedge\dr\Gamma^3\mp q^{\pm 1}\lambda\quint{2}
\Gamma^0\wedge\dr\Gamma^{\pm}\\
&&-q^{\pm 2}\lambda\quint{2}\Gamma^0\wedge\Gamma^3\wedge\Gamma^{\pm}\\
\dr\Gamma^{\pm}\wedge\Gamma^0-\left(1+\lambda^2\right)\Gamma^0\wedge\dr
\Gamma^{\pm}&=&\mp\frac{q^{\mp 1}\lambda}{\quint{2}}\Gamma^3\wedge\dr
\Gamma^{\pm}\pm\frac{q^{\pm 1}\lambda}{\quint{2}}\Gamma^{\pm}\wedge\dr
\Gamma^3\\
&&\pm q^{\pm 1}\lambda^2\Gamma^0\wedge\Gamma^3\wedge\Gamma^{\pm}\\
\dr\Gamma^3\wedge\Gamma^{\pm}-\Gamma^{\pm}\wedge\dr\Gamma^3&=&\pm q^{\mp
1}\lambda\Gamma^3\wedge\dr\Gamma^{\pm}\pm q^{\mp 1}\lambda\quint{2}
\Gamma^0\wedge\dr\Gamma^{\pm}\\
&&+\lambda\quint{2}\Gamma^0\wedge\Gamma^3\wedge\Gamma^{\pm}\\
\dr\Gamma^3\wedge\Gamma^3-\left(1-\lambda^2\right)\Gamma^3\wedge\dr
\Gamma^3&=&\frac{\lambda\quint{2}}{q}\Gamma^+\wedge\dr\Gamma^--\frac{
\lambda\quint{2}}{q}\Gamma^-\wedge\dr\Gamma^+\\
&&-\lambda^2\quint{2}\Gamma^0\wedge\dr\Gamma^3+\frac{\lambda^2
\quint{2}^2}{q}\Gamma^0\wedge\Gamma^-\wedge\Gamma^+\\
\dr\Gamma^3\wedge\Gamma^0-\left(1+\lambda^2\right)\Gamma^0\wedge\dr
\Gamma^3&=&\frac{\lambda}{q}\Gamma^-\wedge\dr\Gamma^+-\frac{\lambda}{q}
\Gamma^+\wedge\dr\Gamma^-+\frac{\lambda^2}{\quint{2}}\Gamma^3\wedge\dr
\Gamma^3\\
&&-\frac{\lambda^2\quint{2}}{q}\Gamma^0\wedge\Gamma^-\wedge\Gamma^+
\end{eqnarray*}
\bigskip
\begin{eqnarray*}
\dr\Gamma^3\wedge\dr\Gamma^{\pm}-q^{\pm 2}\dr\Gamma^{\pm}\wedge\dr\Gamma^3
&=&\pm q^{\pm 1}\lambda\quint{2}\dr\Gamma^0\wedge\dr\Gamma^{\pm}
+q^{\pm 2}\lambda\quint{2}\Gamma^3\wedge\Gamma^{\pm}\wedge\dr\Gamma^0\\
&&-q^{\pm 2}\lambda\quint{2}\Gamma^0\wedge\Gamma^{\pm}\wedge\dr\Gamma^3+
\lambda\quint{2}\Gamma^0\wedge\Gamma^3\wedge\dr\Gamma^{\pm}\\
\dr\Gamma^+\wedge\dr\Gamma^--\dr\Gamma^-\wedge\dr\Gamma^+&=&q\lambda\dr
\Gamma^0\wedge\dr\Gamma^3+\frac{q\lambda}{\quint{2}}\dr\Gamma^3\wedge\dr
\Gamma^3+\lambda\quint{2}\Gamma^0\wedge\Gamma^+\wedge\dr\Gamma^-\\
&&-\lambda\quint{2}\Gamma^0\wedge\Gamma^-\wedge\dr\Gamma^+-\lambda
\quint{2}\Gamma^-\wedge\Gamma^+\wedge\dr\Gamma^0\\
&&-q\lambda^2\Gamma^0\wedge\Gamma^3\wedge\dr\Gamma^3
\end{eqnarray*}
\subsection*{Field Strength Commutation Relations}
\begin{eqnarray*}
F^0\wedge F^a&=&F^a\wedge F^0\\
F^3\wedge F^{\pm}-q^{\pm 2}F^{\pm}\wedge F^3&=&\pm q^{\pm 1}\lambda
\quint{2}F^0\wedge F^{\pm}\\
F^+\wedge F^--F^-\wedge F^+&=&q\lambda F^0\wedge F^3+\frac{q\lambda}{
\quint{2}}F^3\wedge F^3
\end{eqnarray*}

\end{document}